# A Phase Diagram for Crystallization of a Complex Macromolecular Assembly

Vivekananda Bal[≠], Jacqueline M. Wolfrum[§], Paul W. Barone[§], Stacy L. Springs[§],

Anthony J. Sinskey[§ς], Robert M. Kotin[§]*, and Richard D. Braatz[≠§]

[≠]Department of Chemical Engineering, Massachusetts Institute of Technology, Cambridge, MA, USA

[§]Center for Biomedical Innovation, Massachusetts Institute of Technology, Cambridge, MA, USA

[ς]Department of Biology, Massachusetts Institute of Technology, Cambridge, MA, USA

*Gene Therapy Center, University of Massachusetts Chan Medical School, Worcester, MA, USA

**Abstract:** Crystallization of biological molecules has high potential to solve some challenges in drug manufacturing. Thus, understanding the process is critical to efficiently adapting crystallization to biopharmaceutical manufacturing. This article describes phase behavior for the solution crystallization of recombinant adeno-associated virus (rAAV) capsids of serotypes 5, 8, and 9 as model biological macromolecular assemblies. Hanging-drop vapor diffusion experiments are used to determine the combined effects of pH and polyethylene glycol (PEG) and sodium chloride concentrations in which full and empty capsids nucleate and grow. Full and empty capsids show different crystallization behavior although they possess similar capsid structure and similar outer morphology with icosahedral symmetry and 2-fold, 3-fold, and 5-fold symmetry. The differential charge environment surrounding full and empty capsids is found to influence capsid crystallization. The crystal growth rate is found to be affected by the mass of the macromolecular assembly rather than the structure/shape of the macromolecular assembly. The regions of precipitant concentrations and pH in which crystallization occurs are found to be different for different rAAV serotypes and for full and empty capsids for each serotype. Depending on the precipitant concentrations and the rAAV serotype, a variety of complex crystal morphologies are formed and a variety of non-crystallization outcomes such as unidentified dense solid-phase/opaque crystals and an oil/dense phase is observed. The well-defined dense phase/oil is found to be converted into a solid phase over a long period of time. Trends in the crystallization of full and empty capsids between serotypes is observed to be altered by the extent of post-translational modifications (PTMS) associated with the massive macromolecular proteinaceous assembly.

## 1. Introduction

Crystallization of biological molecules, specifically proteins, is becoming increasingly relevant as it has potential for applications in areas including purification, drug formulation, target-specific drug design, long-term storage and transport, and molecular structure analysis (Krauss et al., 2013; Trilisky et al., 2011). Among biological molecules, crystallization of low molecular weight proteins such as lysozyme (~14.3 kDa) is routine, but the crystallization of larger biological macromolecules such as mAbs (~150 kDa), mRNA (200–500 kDa) is not yet well-established. For the crystallization of macromolecular assemblies such as virus capsids (~3.6 MDa), a proteinaceous assembly, only limited resources are available (Ferré-D'Amaré et al., 1998; Lerch et al., 2009; Miller et al., 2006; Mitchell et al., 2023; Rasubala et al., 2005; Saikumar et al., 1998; Trilisky et al., 2011; Xie et al., 2008, 2004, 2002).

This article focuses on the crystallization of adeno-associated virus (AAV) capsids as a model biological macromolecular assembly. AAV capsids are widely used as a vector in gene therapy, because of their relatively low immunogenicity, lack of pathogenicity, high transduction efficiency, lack of off-target genomic alteration, and ability to establish long-term transgene expression (Dunbar et al., 2018; Fattah et al., 2008; Ginn et al., 2018; Ma et al., 2020). There are seven regulatory agency-approved rAAV products for treating rare, monogenic diseases including hemophilia A (Roctavian®), hemophilia B (Hemegenix®), spinal muscular atrophy (Zolgensma®), and Leber's congenital amaurosis type 2 (Luxturna®). In addition, more than 200 rAAV-based gene therapies are in clinical trials (Ginn et al., 2018). The AAV capsid (3.6 MDa) is a complex of 60 protein subunits encoded from the *Cap* gene, VP1 (ca. 87 kDa), VP2 (ca. 72 kDa), and VP3 (ca. 62 kDa) in roughly a 1:1:10 ratio. Stoichiometric expression of the three virion proteins from a single primary transcript results in utilizing alternative splicing and noncanonical translational start codons. The capsids are non-enveloped, icosahedral particles (T=1 symmetry) of approximately 25 nm in diameter with 2-fold, 3-fold, and 5-fold axes of symmetry as shown in our previous work (Bal et al., 2024a; Wörner et al., 2021).

Virions of both the wild-type virus and rAAV contain linear, single-stranded DNA genomes that undergo "rolling hairpin" replication and require only the virus "telomeres" or inverted terminal repeats (ITRs) for replication and encapsidation) (Cotmore and Tattersall, 2013). Vector particles consisting of the capsids and the vector genome (vg) are referred to as *full capsids* and the capsid without vector genome is regarded as *empty capsids* (aka virus-like particles, VLPs). For the full and empty capsids, the calculated pIs for VP3, the major capsid protein, are 5.9 and 6.3, molecular weights are 5.8 MDa and 3.8 MDa, densities are 1.41 and 1.31 gm/cm$^3$ respectively (Kurth et al., 2024; Sommer et al., 2003; Wagner et al., 2023). The molecular weights for full capsids of rAAV5, rAAV8, and rAAV9 are similar: 5.9 MDa, 6 MDa, and 5.85 MDa, respectively.

Previous reports described crystallized rAAV capsids, either empty or full, for each of the serotypes rAAV1 through rAAV9, excluding rAAV5 and rAAV7, but only for few crystallization conditions. In these reports, crystallizations were performed primarily to determine the structure of capsids that would facilitate rational capsid engineering to improve the therapeutic potential of the vectors: for tissue-targeted gene therapy, or to generate viral vectors that can evade the pre-existing antibody responses against the capsid proteins, or to improve therapeutic potency (Kaludov et al., 2003; Lerch et al., 2009; Mikals et al., 2014; Miller et al., 2006; Mitchell et al., 2009; Nam et al., 2011, 2007; Venkatakrishnan et al., 2013; Xie et al., 2008, 2004, 2002). There has not been any effort to date to understand the effect of experimental conditions on the behavior of capsids in solution phase over a broad range of precipitant concentrations (that is, the phase plane/phasic behavior).

In certain applications such as developing a crystallization-based downstream purification method, improving the storage stability and transport properties of gene therapy, or designing drugs with specific bioavailability, dissolution properties, or drug loading, it is important to produce crystals with a specific morphology as it influences many properties of the products such as stability, filterability, drying, mechanical strength, bulk density, and solubility (Mikals et al., 2014; Nam et al., 2007). Thus, it important to explore the crystallization of capsids over a wide range of precipitant concentrations to understand the crystal morphological outcomes and non-crystallization outcomes.

In this work, crystallization conditions are explored over a broad range of precipitant concentrations and the phase diagram is analyzed for both full and empty capsids of rAAV serotypes 5, 8, and 9 at different pH conditions. This will help not only in identifying /designing a successful crystallization condition and obtain crystals of specific morphology for rAAV capsids, but also in understanding the fundamentals of crystallization of complex macromolecular assemblies.

## 2.1 Materials

All chemicals required for the experiments were of molecular biology or cell culture grade. Chemicals purchased from Sigma-Aldrich (St. Louis, MO, USA) were sodium dihydrogen phosphate dihydrate (BioUltra, ≥99.0%), sodium hydroxide (BioXtra, ≥98% anhydrous), 1M hydrochloric acid (BioReagent, for cell culture), sodium chloride (BioXtra, ≥99.5%), potassium chloride (BioXtra, ≥99.5%), potassium dihydrogen phosphate (BioUltra, ≥99.5%), magnesium chloride (BioReagent, ≥ 97%), phosphate buffer saline (PBS 1X (150 mM sodium phosphate and 150 mM NaCl), pH 7.2; BioUltra solution), polyethylene glycol (PEG-8000, BioUltra; PEG-6000, BioUltra), and phosphotungstic acid (10% solution). Poloxamer-188 (Pluronic F-68, 10%, BioReagent) was added to rAAV-containing solutions at 0.001% to suppress binding to container surfaces. Glutaraldehyde solution (10% aqueous solution) and copper grid (carbon support film, 200 mesh, Cu; CF200-CU-50) were purchased from Electron Microscopy Sciences (Hatfield, PA, USA).

## 2.2 rAAV samples

Full (with transgene) and empty (without transgene) capsids of recombinant adeno-associated virus serotypes 5 (rAA5), 8 (rAAV8), and 9 (rAAV9) were obtained from Virovek at a concentration of $10^{14}$ vg/ml. The "full rAAV8" sample actually contains 80% full and 20% empty capsids, and the "empty rAAV8" sample actually contains 96% empty and 4% full capsids.[*] Virovek quantified the full and empty capsids using qPCR and nanodrop OD (optical density) measurement. In-house quantification of full and empty capsids using ddPCR and ELISA produced roughly similar results as reported by Virovek. The full capsids carried a genome of length 2.5kbp (Virovek) with cytomegalovirus (CMV) E1a enhancer promoter and an open reading frame that encodes green fluorescent protein (GFP). Samples were supplied in PBS buffer (pH 7.2). Each vial of 100 μL, was aliquoted into four equal parts and kept at −80°C for long-term storage.

---

[*]Reported by Virovek.

For short-term use, a small vial was thawed, and stored at 4°C, at which AAV is stable for 4 weeks.

## 2.3 Experiment

Hanging drop vapor diffusion experiments (**Fig.1**) were carried out in VDX 24-well crystallization plates (Hampton Research, California, USA) to screen crystallization conditions (Miller et al., 2006; Nam et al., 2011). In a hanging drop experiment, each well contains a droplet (2 μL) hanging from a glass cover slip at the top of the well, and reservoir solution (1 mL) at the bottom of the well. Both the drop and reservoir solution contain sodium chloride (NaCl) and polyethylene glycol (PEG) 8000 precipitants in phosphate-buffered saline (PBS). In the enclosed well, the reservoir ensures the transfer of water from the droplet to the reservoir and consequent building towards supersaturation in the droplet that attains quasi-steady conditions (**Fig. 1**).

PEG is one of the most commonly used precipitants for protein crystallization because of its well-known local solubility reduction by the so-called *volume exclusion effect* (Bonneté, 2007; Charles et al., 2006; Weber et al., 2008). PEG8000 has been used in rAAV capsid crystallization, where it has been observed that, as the molecular weight of the PEG increases, lower quantities of capsids are needed to achieve local supersaturation concentrations (Kaludov et al., 2003; Lerch et al., 2009; Miller et al., 2006; Nam et al., 2011, 2007; Xie et al., 2008, 2004, 2002). PEG8000 was used, as in past studies, at an initial concentration of about 2 to 7% wt:vol, to nucleate rAAV capsid crystals in a variety of buffer solutions.

Each droplet contained rAAV capsids in a PBS (1 μL) sample and a reservoir solution (1 μL). The initial concentration of PEG and NaCl in the droplet was half of that in the reservoir solution. Due to vapor pressure depression, the water vapor diffuses from the droplet to the reservoir increasing the protein concentration in the droplet. Once supersaturated conditions are reached and sufficient time passes, capsid particles nucleate and grow. Eventually, the evaporation rate slows and a quasi-steady state is approached. Scouting crystallization conditions was performed by varying the concentration of pH, PEG8000, and NaCl concentrations.

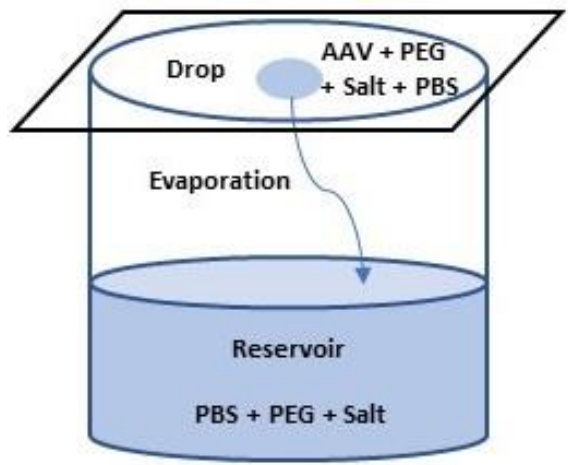

**Fig. 1:** Hanging-drop vapor diffusion experimental setup.

Optical microscopy was used to monitor the evolution of crystals (Imaging Source DMK42BUC03) at regular time intervals for a period of 1 to 2 weeks. The particles were observed under a cross-polarized light microscope (Leica Z16 APO) to confirm the crystallinity of the particles. Crystal growth rates were found to slow down within a few days of nuclei formation. The pH range was limited to a range between 5.5 and 8.5(Gruntman et al., 2015; Lins-Austin et al., 2020), as capsid proteins are more stable within this pH range. All the experiments were carried out at room temperature, which was 23±2°C.

### 2.3.1 Crystallinity confirmation

Given the number of crystallization conditions, it is not practical to use either X-ray diffraction (XRD) or scanning electron microscopy (SEM) imaging to screen crystals in each droplet. Therefore, particle crystallinity was established by the birefringence property of anisotropic crystals under cross-polarized light as described in detail in our previous work (Bal et al., 2024b). Screening of crystallization conditions using cross-polarized light is simpler and faster than either XRD or SEM imaging. Micron-sized crystals are readily observed using optical microscopy. Isotropic crystals are not birefringent under cross-polarized light and non-crystalline precipitates will also not show birefringence. Some dense/opaque crystals are not birefringent and remain dark under cross-polarized light. It is difficult to characterize the crystallinity of those crystals/particles; therefore, those crystallization conditions were excluded in the analysis. Here we provide a brief description of the screening using cross-polarized light microscopy.

For the polarization study, a 24-well plate was positioned on a Leica Z16 APO microscope base. Then plane-polarized light is passed through the crystals and the analyzer is rotated 360° while keeping the crystal's position fixed. On rotation of the analyzer, anisotropic crystals assume a spectrum of interference color at all angles except at every 90° position of the analyzer, where the interference is extinct and the crystals appear dark. This birefringence confirms the crystallinity of particles.

### 2.3.2 Construction of phase diagram

To construct a phase diagram, over 500 crystallization conditions are screened by varying PEG and NaCl concentrations at a specific pH. The outcome of every experimental condition is noted and the experimental outcome and corresponding pH condition are plotted as a function of precipitant concentrations to obtain a phase diagram for that pH. Each data point in a phase diagram carries both vertical and horizontal error bars from three repeated runs.

## 3 Results and Discussion

This section reports and discusses the phase diagrams for the crystallization of rAAV capsids. Crystallization conditions are explored for both full and empty capsids of the rAAV5, rAAV8 and rAAV9 serotypes. First, the concentration of the rAAV over which crystallization occurs is studied and then the phase diagram analysis is performed.

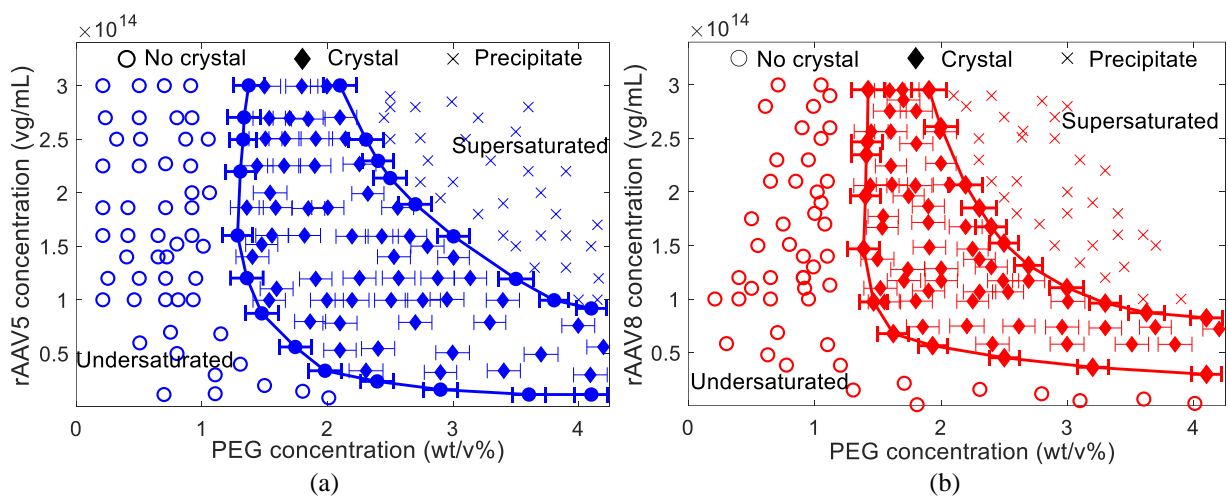

**Fig. 2:** Crystallization/precipitation conditions for full (a) rAAV5 and (b) rAAV8 capsids as a function of initial PEG8000 concentration in droplets for NaCl concentration of 1 M at pH 7.4. The experimental results for empty capsids are within the error bars shown for full capsids.

PEG is one of the most commonly used precipitants for protein crystallization because of its well-known local solubility reduction by the so-called *volume exclusion effect* (Bonneté, 2007; Charles et al., 2006; Weber et al., 2008). PEG8000 has been used in rAAV capsid crystallization, where it has been observed that, as the molecular weight of the PEG increases, lower quantities of capsids are needed to achieve local supersaturation concentrations (Kaludov et al., 2003; Lerch et al., 2009; Miller et al., 2006; Nam et al., 2011, 2007; Xie et al., 2008, 2004, 2002). PEG8000 was used, as in past studies, at an initial concentration of about 2 to 7% wt:vol, to nucleate rAAV capsid crystals in a variety of buffer solutions.

A wide range (0.5% to 8% wt:v) of PEG8000 concentrations was explored to scout conditions that nucleate crystals of full and empty AAV capsids (**Fig. 2**). rAAV5 crystals nucleated at lower initial PEG8000 concentrations than in previous reports (Lerch et al., 2009; Sommer et al., 2003; Xie et al., 2008, 2004, 2002), which may be attributed to the use of a different buffers in the studies. The range of PEG8000 concentration over which crystallization occurs (**Fig. 2**) is somewhat wider for rAAV5 (**Fig. 2a**) than for rAAV8 (**Fig. 2b**) and that reported for rAAV2 (Xie et al., 2004). Similar to rAAV2, the width of the metastable zone narrows as the concentration of rAAV capsids increases. As the rAAV concentration increases, the range of PEG over which crystallization occurs decreases. Below we carry out phase diagram analysis at rAAV concentration of $1 \times 10^{14}$ vg/mL, where the crystallization window for the PEG is wide.

The inorganic precipitant, NaCl, influences the crystallization process by shifting the protein solubility whereas the organic precipitant, PEG, functions as an aquacide effectively concentrating the proteins. Sodium chloride is widely used as a precipitant in the crystallization of proteins(Bonneté, 2007; Charles et al., 2006; Hekmat, 2015; Weber et al., 2008; Yamanaka et al., 2011) to shift the protein solubility by changing the number of water molecules available for interaction with the charged part of a protein's surface.[†] While NaCl was added in previous AAV capsid crystallization studies (Lerch et al., 2009; Mitchell et al., 2009; Nam et al., 2011; Xie et al., 2008, 2004, 2002), this study is the first to quantify the effect of the *variation* in NaCl concentration on the crystallization of AAVs.

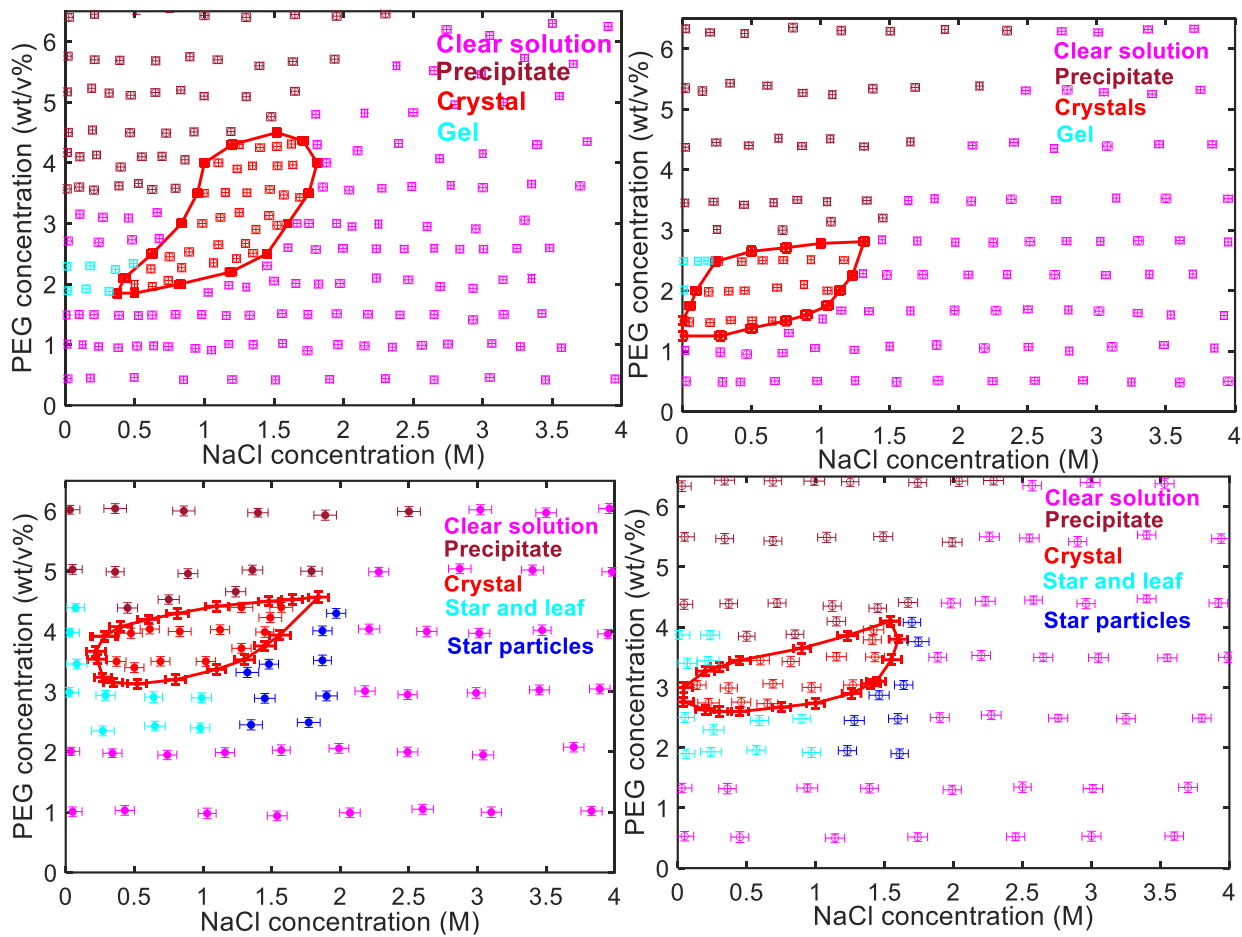

**Fig. 3:** Phase diagrams for the crystallization of rAAV capsids. Row 1: full rAAV5 capsids (left) and empty rAAV5 capsids (right). Row 2: full rAAV9 capsids (left) and empty rAAV9 capsids (right) at pH 5.7. The phase diagrams show the regions favorable for crystallization, precipitate formation, gel formation, and undersaturated solution as a function of PEG8000 and NaCl. Each experimental data point is the average of the three experimental repetitions.

**Fig. 3** shows the experimental data points with the corresponding experimental outcome for full (left) and empty (right) capsids of rAAV5 (row 1) and rAAV9 (row 2) as a function of NaCl and PEG800 concentrations at pH 5.7. Experimental data points with the similar outcome are labelled with the same color. The regions favorable for crystallization, precipitate formation, gel formation, and undersaturated (clear) solution as a function of PEG8000 and NaCl at a specific pH are known in the phase diagrams. The largest difference between full and empty capsids is that crystallization of empty rAAV9 capsids can occur at much lower NaCl concentrations than for full rAAV9 capsids.

---

[†] A reduction in protein solubility (aka *salting out*) is associated with a decrease in the number of water molecules, whereas an increase in protein solubility (aka *salting in*) is associated with an increase in the number of water molecules (Chernov, 2003; Grover and Ryall, 2005; Hyde et al., 2017; McPherson and Gavira, 2014).

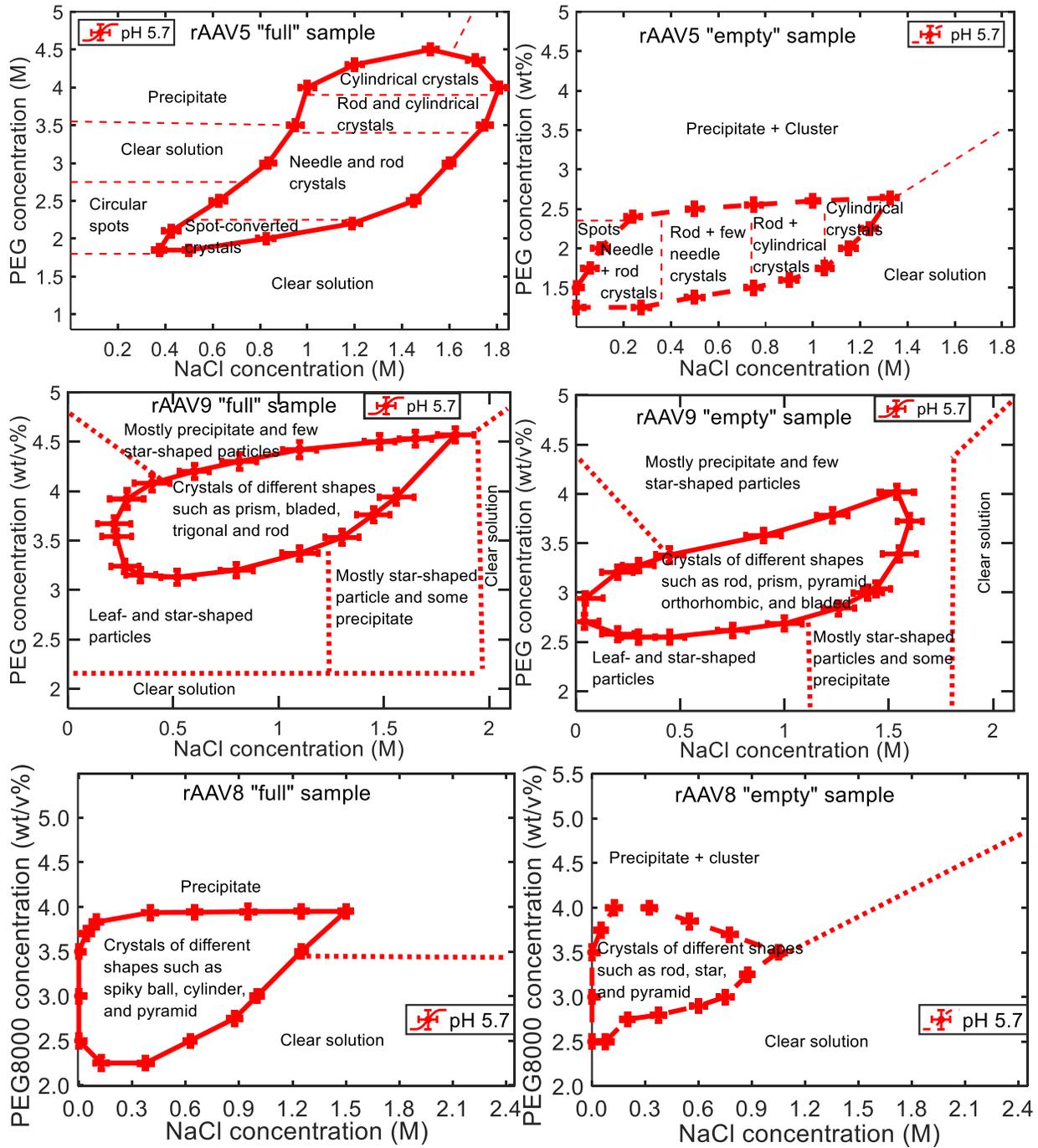

**Fig. 4:** Phase diagrams for the crystallization of rAAV capsids showing the different morphological outcomes as a function of PEG8000 and NaCl concentrations. Row 1: full rAAV5 (left) and empty rAAV5 (right). Row 2: full rAAV9 (left) and empty rAAV9 (right). Row 3: full rAAV8 (left) and empty rAAV8 (right). The initial capsid concentration is $10^{14}$ vg/ml and pH is 5.7.

**Fig. 4** shows the phase diagrams for full and empty capsids providing a higher level, zoomed-in map of crystallization conditions with morphological outcomes over a wide range of precipitant concentrations at pH 5.7. For each capsid type, crystallization occurs within a closed and bounded region. Full and empty capsids show significantly different phase behaviour at different PEG and NaCl concentrations. While AAV capsids have been crystallized previously, past efforts have not included detailed analysis of the phase behavior (Lerch et al., 2009; Mitchell et al., 2009; Nam et al., 2011; Xie et al., 2008, 2004, 2002). Based on PEG and NaCl concentrations, a variety of non-crystallization outcomes can occur including forming a dispersed phase (gel spots as in **Fig. 4 row 1**), undefined cluster-like spots (as in **Fig. 4 rows 1 and 3**), precipitates (**Fig. 4 rows 1–3**), and clear solution, which are discussed later in this section. The solution remains clear at high NaCl concentration and low to intermediate PEG concentration (**Fig. 4 rows 1–3**).

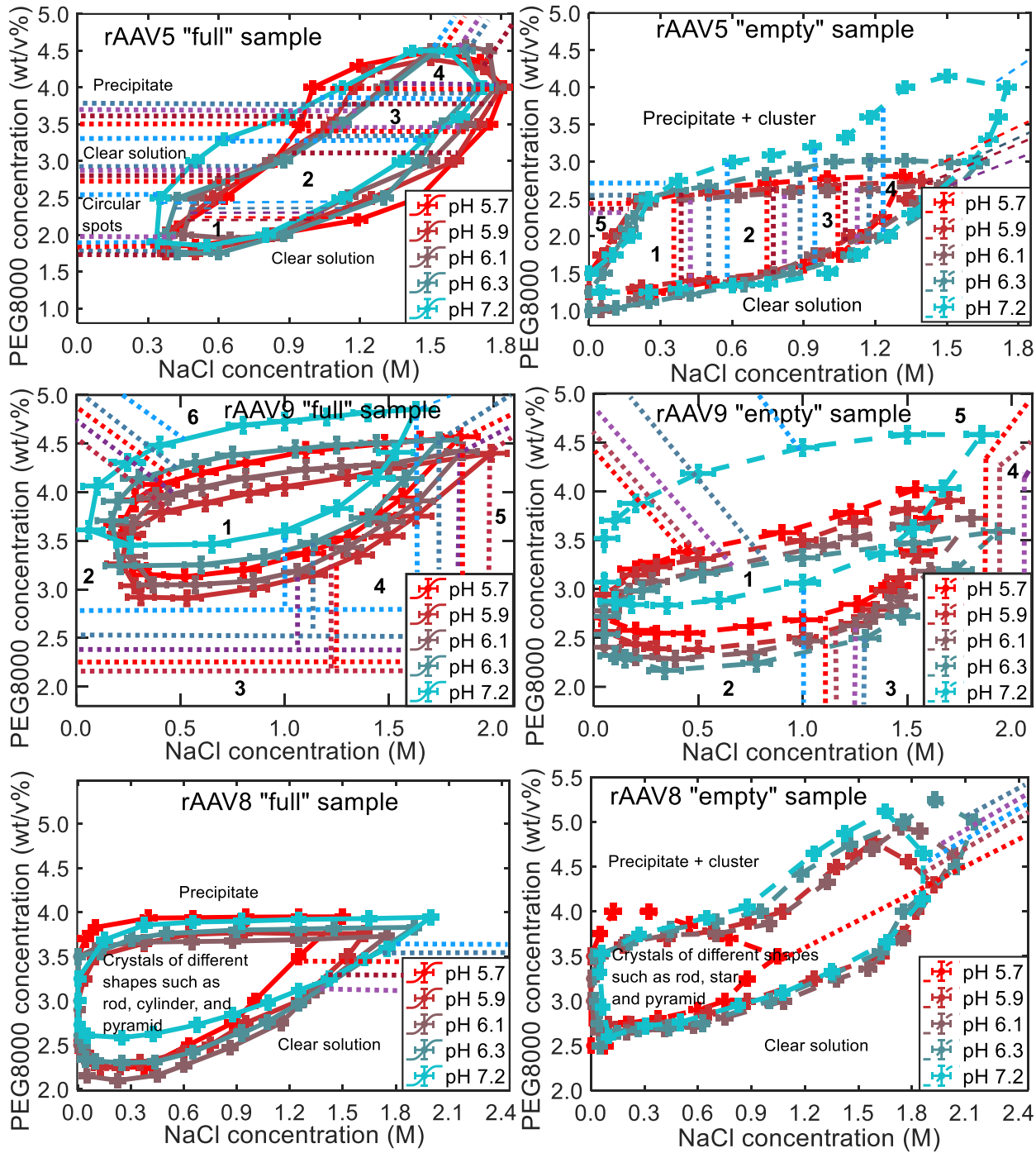

**Fig. 5:** Phase diagrams for the crystallization of rAAV capsids as a function of pH. Row 1: full (left) and empty (right) capsids of rAAV5. For full rAAV5: 1 = spot (gel/oil) converted crystals; 2 = needle and rod crystals; 3 = rod and cylindrical crystals; 4 = cylindrical crystals. For empty rAAV5: 1 = needle and rod crystals; 2 = rod and few needle

crystals; 3 = rod and cylindrical crystals; 4 = cylindrical crystals; 5 = circular gel spots. Row 2: full (left) and empty (right) capsids of rAAV9. For full rAAV9: 1 = crystals of different shapes such as prism, bladed, trigonal, and rod; 2 = leaf- and star-shaped particles; 3 = clear solution; 4 = mostly star-shaped particles and some precipitates; 5 = clear solution; 6 = mostly precipitates and few star-shaped particles. For empty rAAV9: 1 = crystals of different shapes such as rod, prism, orthorhombic, bladed, needle ball, and pyramid; 2 = leaf- and star-shaped particles; 3 = mostly star-shaped particles and some precipitates; 4 = clear solution; 5 = mostly precipitates and few star-shaped particles. Row 3: full (left) and empty capsids of rAA8. The initial capsid concentration is $10^{14}$ vg/ml.

**Fig. 5** shows the phase diagrams as a function of pH. For full AAV5 capsids (**Figs. 4** and **5 row 1 left**), the crystallization zone is narrowest at the low PEG concentration and is widest for intermediate PEG concentration, and the width of crystallization zone varies with both PEG and pH. This variation in the width of the crystallization zone is largest at low pH (5.7). At PEG concentration less than ~1.8 wt/v% for any NaCl concentration, the solution remains clear. At low NaCl concentration (e.g., < 0.3 M), for PEG concentration ~2–2.5 wt/v%, circular spots form (indicating a phase separation). On further increase in PEG concentration (e.g., to ~3 wt/v%), the solution remains clear and, at even higher PEG concentration, precipitation is observed. The formation of spots and clear solution are explained later in this section. A variety of crystal morphologies are observed for full rAAV5 capsids. Intermediate PEG and NaCl concentrations (e.g., 3 wt/v% PEG and 1 M NaCl) favor the formation of needle-shaped crystals, whereas higher PEG concentration (e.g., 3.7 wt/v% PEG) favors the formation of rod-like and cylindrical crystals.

For empty AAV5 capsids, the range of NaCl concentration over which crystallization occurs is shifted to lower values (**Figs. 4** and **5 row 1 right**). As for full rAAV5 capsids, the crystallization zone shifts to a higher NaCl concentration range as the PEG concentration increases. The overall area (in the NaCl-PEG plane) over which the crystallization of capsids occurs increases proportionally with pH for empty capsids, which is not significant for full rAAV5 (**Figs. 4** and **5 row 1 left**). Empty AAV5 capsids crystallize over a narrow range of PEG concentration, at about 1.2 wt/v%, without addition of salt, whereas full rAAV5 capsids fail to form crystals in the absence of NaCl. As for full rAAV5, the solution remains clear at low PEG concentration and, at intermediate PEG and low NaCl concentration, spots appear (**Figs. 4** and **5 row 1 right**). At high PEG concentration, AAV5 precipitates form. Empty rAAV5 capsids mostly form needle-shaped crystals at low NaCl concentration and, as the NaCl concentration increases, transitions to rod-like and then to cylindrical crystals.

The crystallization zone for full rAAV9 is narrow at high PEG concentrations and the widest at low PEG concentration (**Figs. 4** and **5 row 2 left**). This variation in the width of the crystallization zone is largest at pH 5.9. A PEG concentration less than ~2.75 wt/v% and for NaCl concentration less than ~1.8M, star- and leaf-shaped particles formed; precipitation occurs at PEG concentrations higher than ~4.75 wt/v% and for any NaCl concentration; the solution remains clear at PEG concentrations below ~1.5 wt/v% for any NaCl concentration and at NaCl concentrations greater than ~2M for any PEG concentration. As for rAAV5 (**Figs. 4** and **5 row 1**), the full rAAV9 crystallization zone moves to the higher NaCl concentration as the PEG concentration increases. In contrast to rAAV5, full rAAV9 mostly forms complex crystal morphologies such as rod, cylinder, needle, orthorhombic, prism, and bladed (**Figs. 4** and **5 row 2 left**), and the full range of morphologies occur throughout the crystallization range for full rAAV9 capsids.

For empty rAAV9 (**Figs. 4** and **5 row 2 right**), the crystallization zone is the widest for intermediate PEG concentration and narrowest for high PEG concentration. As with empty rAAV5, crystallization can occur at higher NaCl and PEG concentrations for some values of pH. The area of the region where crystallization occurs is smallest at low pH (5.7) and largest at pH 7.2 as observed in **Fig. 5 row 2 right**. With the full rAAV9 (**Fig. 5 row 2 left**), the solution remains clear at higher NaCl concentration and at lower PEG concentration, precipitate forms at high PEG concentration, and star and leaf-shaped particles are formed at low NaCl and intermediate PEG concentrations (**Fig. 5 row 2 right**). Like full rAAV9, empty rAAV9 mostly forms complex crystal morphologies such as bladed, prism, pyramid, and orthorhombic. These crystal morphologies can occur anywhere within the crystallization region, which are not observed in either full or empty rAAV5 crystals. Much higher PEG concentration is required to cause the crystallization of empty rAAV9 than for empty rAAV5 (cf. **Figs. 5 row 1** and **row 2 right**).

Both full and empty rAAV9 capsids do not form gel spots (**Figs. 5 row 2**), which are observed in both full and empty rAAV5 (**Figs. 5 row 1**). Unlike with full rAAV5 (**Fig. 5 row 1 left**), crystallization of both full and empty rAAV9 does not occur without the addition of salt (**Figs. 5 row 2**). Star- and leaf-shaped particles formed in both full and empty rAAV9 are most likely dense/opaque crystals which do not show any birefringence under cross-polarized light and remain dark. However, lacking proper characterization techniques and difficulty in separating the complex crystals made it impossible to analyze the particles further. Thus, experimental conditions favouring the formation of star- and leaf-shaped particles are excluded from analysis.

For full rAAV8 capsid (**Figs. 4** and **5 row 3 left**), the range of NaCl concentration over which crystallization occurs is narrow at low PEG concentration and the widest at high PEG concentration. As observed in **Fig. 5 row 3 left**, this variation in crystallization zone width is largest at high pH (7.2), which is opposite to that obtained for full rAAV5 and rAAV9 crystallization, for which largest variation in the width of crystallization zone occurs at low pH. At PEG concentration less than ~2 wt/v% and for any NaCl concentration, the solution remains clear; precipitation occurs at PEG concentration higher than ~4.1 wt/v% and for any NaCl concentrations. Similar to full and empty rAAV5 capsids (**Figs. 5 row 1**), a variety of crystal morphologies such as rod, cylinder, and pyramid are observed in the crystallization region (**Figs. 4** and **5 row 3 left**), with a difference being that the full range of morphologies occur throughout the crystallization range for full rAAV8 capsids. Formation of full range of morphologies throughout the crystallization range also observed for rAAV9 capsids.

For the empty AAV8 capsids (**Figs. 4** and **5 row 3 right**), the crystallization zone is the widest for intermediate PEG concentration and narrowest for high PEG concentration. As for the empty AAV5 capsids, crystallization can occur at higher NaCl and PEG concentrations at several pH conditions. The area of the region where crystallization occurs is smallest at low pH (5.7) and largest at intermediate pH. This drastic variation in area of crystallization region with pH is not observed for either full or empty capsids of AAV5 and AAV9. As for the other capsids, the solution remains clear at low PEG concentration and precipitates form at high PEG concentration (**Fig. 5 row 3 right**). Like both full and empty rAAV9 (**Figs. 5 row 2**), empty rAAV8 (**Fig. 5 row 3 right**) mostly forms complex crystal morphologies such as star, prism, and pyramid, rod, which are not observed with either full or empty AAV5 capsids (**Figs. 5 row 1**). As for full rAAV8 and for both full and empty AAV9 capsids, these crystal morphologies appear anywhere within the crystallization region. Much higher PEG concentration is required to cause the crystallization of empty rAAV8 than for empty rAAV5 (cf. **Figs. 5 rows 1 and 3**).

Like both full and empty AAV9 capsids (**Figs. 5 row 2**), full and empty AAV8 capsids (**Figs. 5 row 3**) do not form gel spots, which are observed in both full and empty AAV5 capsids (**Figs. 5 row 1**). As for empty AAV5 capsids (**Fig. 5 row 1 right**), there is a narrow range of PEG concentration where crystallization of either full or empty AAV8 capsids is possible without the addition of salt, which does not occur for full and empty rAAV9 capsids.

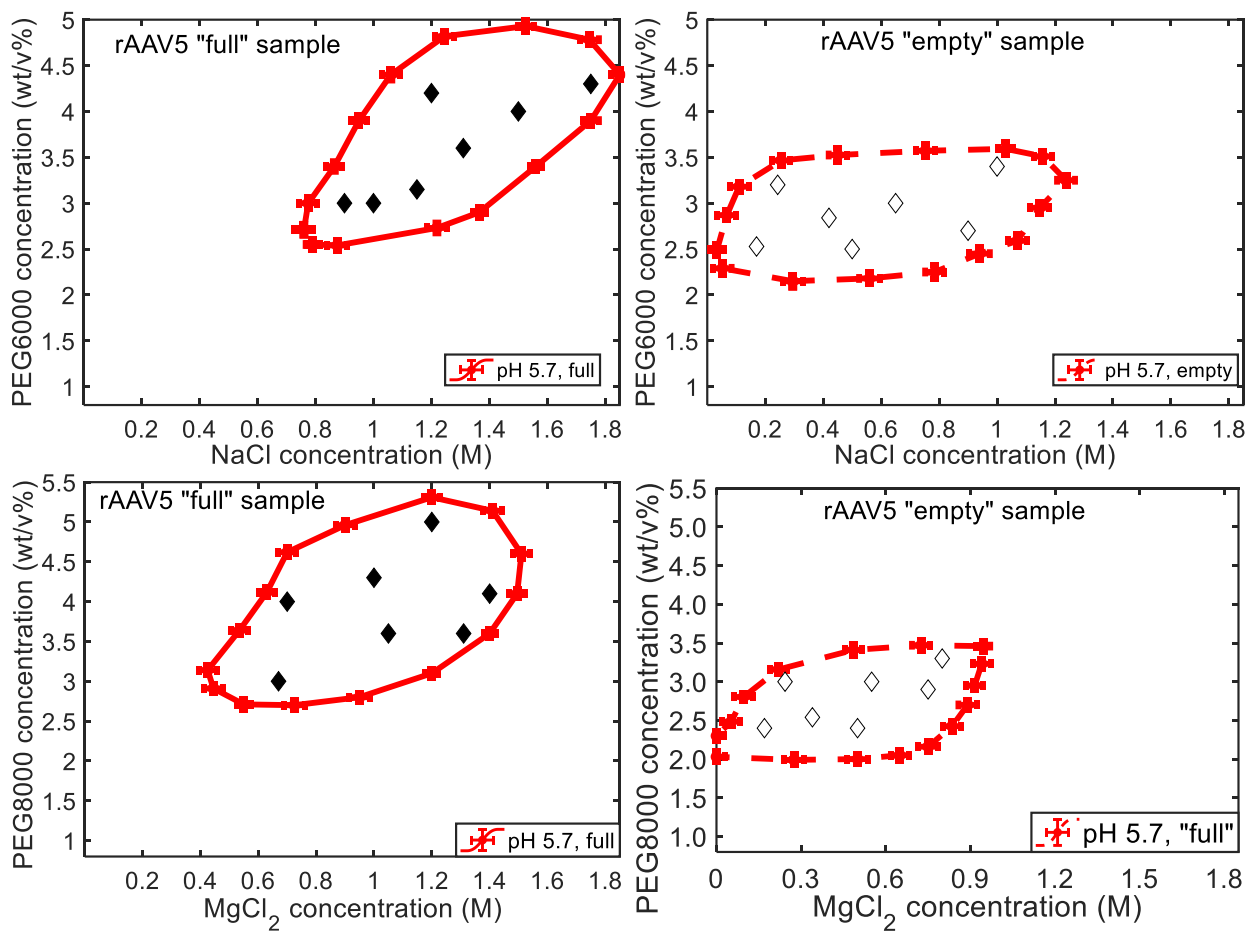

**Fig. 6:** Effect of molecular weight of PEG and MgCl$_2$ on the crystallization phase diagram. Effect of PEG8000 on the crystallization phase diagram for full capsids (**row 1 left**) and empty capsids (**row 1 right**) and effect of MgCl$_2$ on the crystallization phase diagram for full capsids (**row 2 left**) and empty capsids (**row 2 right**). The initial capsid concentration is $10^{14}$ vg/ml.

All the previous screenings of the crystallization conditions used PEG8000 and NaCl as precipitants. In order to understand whether the method of preferential crystallization works with other precipitants, crystallization conditions were screened using PEG6000 and NaCl combination (**Fig. 6 row 1**), and PEG8000 and MgCl$_2$ combination (**Fig. 6 row 2**) as precipitants. Similar to the PEG8000 and NaCl combination (**Fig. 5**), in case of both PEG6000 and NaCl combination, and PEG8000 and MgCl$_2$ combination, shape of phase diagrams for crystallization regions for both full and empty capsids remain nearly the same. In contrast to the PEG8000 and NaCl combination, in case of PEG6000 and NaCl combination (**Fig. 6 row 1**), crystallization region lies at relatively higher PEG concentration to build the supersaturation level favourable for crystallization due to the relatively low volume exclusion effect of low molecular weight PEG. Likewise, in case of MgCl$_2$, crystallization region occurs at relatively lower salt concentration than in case of NaCl.

This is because higher ionic strength associated with MgCl$_2$ needs lower salt concentration to build the supersaturation level conducive for the crystallization. Thus, preferential crystallization works well with PEG of any molecular weight and with inorganic salt of any valency.

We carried out an analysis (Supplementary material **Fig. S1**) to rule out the possibility that any of the particles formed in this study consist of NaCl or PEG. The NaCl concentrations used in the study ranged between 0.0 to 2.5 M and are much lower than the solubility of NaCl (~6 M) in aqueous solution at room temperature (Bharmoria et al., 2012; Gordon & Ford, 1972). Similarly, the PEG concentrations explored in this article of 0.5 to 8 wt/v% are much lower than the solubility of PEG (~63 wt/v%, sigmaaldrich.com) in aqueous solution at room temperature. The literature suggests that, among these two precipitants, the presence of one precipitant can reduce the solubility of another in aqueous

solution,(Brunchi and Ghimici, 2013; Heeb et al., 2009; McGarvey and Hoffmann, 2018; Soleymani et al., 2020), so we carried out a set of control experiments for the combi-nations of the full ranges of PEG concentration (0.5 to 8 wt/v%) and NaCl concentration (0.005 to 2.5 M) in the absence of rAAV capsids. No crystals formed in these control experiments, which supports the conclusion that the crystals formed in our experiments in presence of rAAV are of capsids.

To verify crystallinity of particles, they were observed through normal (Imaging Source DMK42BUC03) and cross-polarized light microscope (Leica Z16 APO). As expected, crystals appear grey in normal microscope and bright, colorful in cross-polarized light microscope as reported in our previous work (Bal et al., 2024). Due to the lack of proper characterization tools, particles/crystals, which do not show birefringence, are excluded from the phase diagram construction. Star- and leaf-shaped particles formed in both full and empty rAAV9 are most likely opaque crystals, which do not show any birefringence under cross-polarized light and remains dark. Thus, the crystallization conditions inside the phase diagram produce crystals, which are optically transparent and anisotropic.

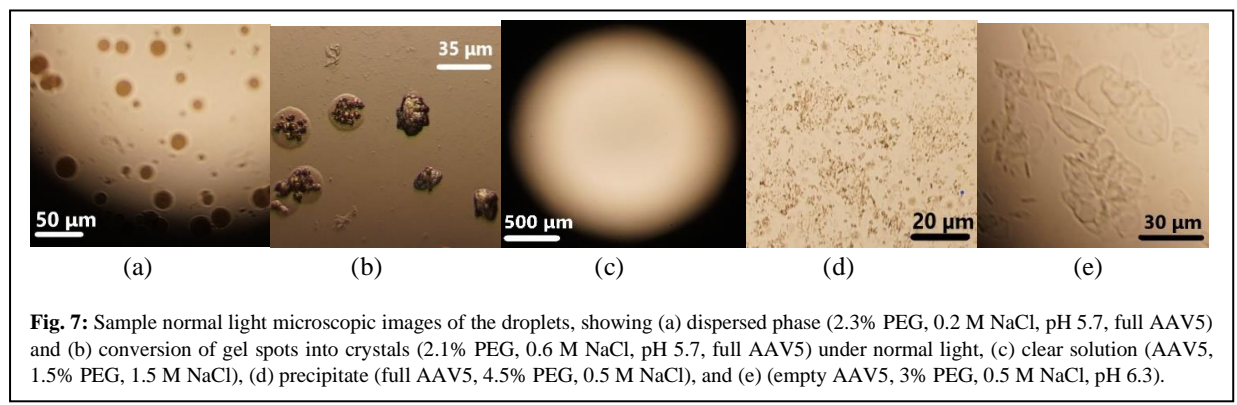

**Fig. 7:** Sample normal light microscopic images of the droplets, showing (a) dispersed phase (2.3% PEG, 0.2 M NaCl, pH 5.7, full AAV5) and (b) conversion of gel spots into crystals (2.1% PEG, 0.6 M NaCl, pH 5.7, full AAV5) under normal light, (c) clear solution (AAV5, 1.5% PEG, 1.5 M NaCl), (d) precipitate (full AAV5, 4.5% PEG, 0.5 M NaCl), and (e) (empty AAV5, 3% PEG, 0.5 M NaCl, pH 6.3).

At lower PEG concentration, the solution at all NaCl concentrations for full and empty capsids for all serotypes was expected to be clear due to the low volume exclusion effect of PEG and consequent undersaturation inadequate to form crystals. At intermediate PEG and low salt concentrations for rAAV5, the formation of spots was observed (**Figs. 4** and **5 row 1**). The composition of the spots is attributed to a gel/oil phase, which is a protein-rich phase (**Fig. 7a**). Crystals form in some of these spots (**Fig. 7b**) after an extended time (e.g., 7 to 14 days) and other spots remain unchanged. Although not previously reported in the literature on capsid crystallization, the formation of such spots at low NaCl concentration (at low ionic strength) has been reported in the broader protein crystallization literature (Asherie, 2004; Broide et al., 1991; Martin et al., 2021; Ozbas et al., 2004; Pegram et al., 2010; Zayas, 1997; Ziegler and Foegeding, 1990). These observations are consistent with reports of protein solubility initially increasing monotonically with increasing NaCl concentration ("salting in") until reaching a maximum solubility at an ionic strength of ~2.9 to 4 M (~2.6–3.8 M NaCl) (Mehta et al., 2012; Quigley and Williams, 2015a). An increase in protein solubility at lower salt concentration agrees with calculations of the protein-protein potential of mean force in terms of second virial coefficients for different proteins in the presence of different salts, which corresponds with higher protein solubility and decreasing attractive protein-protein interactions (Curtis and Lue, 2006; Dumetz et al., 2007; Haas et al., 1999; Mehta et al., 2012; Quigley and Williams, 2015). For the NaCl concentrations explored in this study, the ionic strength remains well below the maximum solubility limit found in the literature for NaCl (Arakawa and Timasheff, 1985; Arawaka et al., 1991; McPherson and Gavira, 2014; Zhang et al., 2012). For NaCl concentrations below this limit, the low solubility results in a high supersaturation, and the formulation of a new protein-rich phase (dispersed phase) known as a gel/oil. The high rate does not provide enough time for the capsids to orient into a crystal lattice to form a crystal, hence a gel forms.

At high PEG concentration (e.g., 5 wt/v%), for all serotypes and all values of NaCl concentration, very high supersaturation is associated with the very low solubility and presumably leads to the formation of precipitate before the protein-rich gel/oil phase can form. Precipitation occurs at high PEG concentrations for both full and empty capsids of rAAV5, rAAV8, and rAAV9. At high NaCl concentration at lower PEG concentration, the high capsid solubility is associated with undersaturation or sub supersaturation, so no new phase forms during the experiment and the solution remains clear with no precipitate (**Figs. 4** and **5**). The lack of particle formulation under these conditions is consistent with past studies that have shown that the protein solubility increases even at high values of NaCl concentration (Chirag M Mehta et al., 2012; Quigley and Williams, 2015a). Previous reports on rAAV also described the absence of crystals at higher NaCl concentration (as high as 1 M); (Lerch et al., 2009; Mitchell et al., 2009), however, the references do not provide a detailed analysis of conditions and formation.

Within the crystallization region (**Figs. 4** and **5**), the NaCl concentration remains within the soluble range, which results in supersaturated conditions favorable for crystallization. Moderate supersaturation is associated with a moderate driving force (i.e., the difference in concentration of capsids between supersaturation state and saturation state), which reduces the rate of formation of a new solid phase, giving the capsids sufficient time to orient to form and grow as crystals. The solution conditions identified for the formation of capsid crystals are consistent with past studies for empty AAV9 capsids (all in different buffer though).(Halder et al., 2015). These results are the first report for the crystallization of full capsids of rAAV5, rAAV8, and rAAV9. For all the serotypes, the range of PEG concentration in which crystallization occurs shifts to higher values with higher NaCl concentration for most values of pH (**Fig. 5**). This positive correlation between PEG and NaCl concentrations that induce crystal formation is likely associated with the precipitants' opposing effects on the protein solubility over the experimental conditions in this study. As PEG concentration increases, the capsid solubility decreases. On the other hand, as NaCl concentration increases, the capsid solubility increases under the experimental conditions. These competing effects would mostly cancel, so that the supersaturation remains at intermediate values conducive to crystallization (Asherie, 2004; McPherson and Gavira, 2014; Randolph, A. D.; Larson, 1971).

The sensitivity of the crystallization regions on pH is stronger for empty than for full capsids for some values of pH (**Fig. 5**). A potential cause for this difference in dependency is that the isoelectronic point of 6.3 for empty capsids is clearly in the center of the pH range in the experiments, and of 5.9 for full capsids is near the edge of the pH range. The largest change in charge distribution/zeta potential occurs when the pH crosses the isoelectronic point, which is fully spanned for the empty capsids. Thus, on moving from pH 6.3 to either 7.2 or 6.1, largest change in capsid charge/zeta potential and in the resulting solubility occurs for empty capsid than for full capsids (Shaw et al., 2001). This is observed by the strong variation of crystallization region with precipitants, which is reflected by the strong sensitivity of crystallization region to the solution pH for empty capsids.

That capsid crystallization occurs at values of pH that are in the vicinity of the isoelectric point is consistent with past studies on protein crystallization and precipitation. The capsid solubility is at a minimum for pH near the isoelectric point, as favorable interactions between protein and solvent is minimal due to net zero charge on capsids (Jiang et al., 2010; Schmittschmitt and Scholtz, 2003; Shaw et al., 2001). Also, the electrostatic repulsion between proteins is the lowest when the proteins are at net zero charge. The same result holds for other molecules in solution and would also hold for capsids. For full and empty capsids (**Fig. 5**), crystallization at higher pH is enabled at higher PEG by increasing the NaCl concentration. This observation is consistent with the higher ionic strength associated with higher NaCl concentration neutralizing the repulsive charges of the capsids allowing individual capsids to physically associate despite capsid surfaces that are oppositely charged locally.

Although the shape of crystallization regions is qualitatively similar for all the serotypes for both full and empty capsids (**Fig. 5**), dependence of empty rAAV5 (**Fig. 5 row 1 left**) on pH seems to be different from others. For all capsids, increasing pH from 5.7 to pI (which is 5.9 for full capsids and 6.3 for empty capsids), there is an increase in the size (i.e., shift) of the crystallization region towards relatively higher NaCl concentration. This is because decreasing solubility with increasing pH within this pH range needs higher ionic strength to maintain supersaturation conducive to crystallization. However, further increasing pH from pI to 7.2, the crystallization regions for all capsids except empty AAV5 shifts to relatively lower NaCl concentration. This is because for all the capsids except empty AAV5, crystallization region reached the ionic strength close to maximum solubility (~ 2 M NaCl (Chirag M Mehta et al., 2012)) and thus on further increasing pH to 7.2 shifts the crystallization region to relatively lower NaCl. On the contrary, for empty AAV5, the NaCl concentration used is still below the maximum solubility limit, which allows further addition of NaCl. However, this change in the size of the crystallization region is much more significant for empty capsids than full capsids.

**Fig. 5** suggests that, for rAAV5 and rAAV9, full capsids crystallize at a PEG concentration relatively higher than that for empty capsids. This trend is reversed for rAAV8. This effect may be partially attributed to the difference in level of post-translational modifications (PTMS) between full and empty capsids for different serotypes (Mary et al., 2019). Higher PTMS level (Supplementary material **Fig. S2**) in empty capsids compared to that in full capsids results in higher charge density and resulting higher solubility, which is reflected by crystallization of empty capsids at higher PEG concentration. In contrary, for rAAV5 and rAAV9, relatively lower PTMS level in empty capsids compared to that in full capsids causes the former to crystallize at lower PEG concentration.

## 4 Conclusion

This work maps out the full range of crystallization conditions for capsids of three rAAV serotypes (5, 8, and 9) as model biological macromolecular assemblies and analyses the phase diagram. In contrast to the low molecular weight protein crystallization and like large macromolecular crystallization (Ahamed et al., 2007; Iwai et al., 2008; Liu et al., 2010; Rakel et al., 2014; Sedgwick et al., n.d.; Trilisky et al., 2011), capsid crystallization is found to occur within a narrow region, which shows an irregular and nonlinear variation with pH, precipitant concentrations and serotypes, but in contrast to the other proteins, crystallization regions are found to be bounded. Like other macromolecular proteins such as mAbs (Ahamed et al., 2007; Rakel et al., 2014; Trilisky et al., 2011), a range of complex crystal morphologies such as prisms, trigonals, whiskers, spikey balls, and undetermined shapes are formed for capsids depending on the crystallization conditions and the serotypes. The formation of much more complex/intricate morphologies such as leaves, 2D and 3D stars, squid heads, sea urchins, and dendrites for capsids; however, suggests that its solution chemistry is different from that of other macromolecular crystallization leading to the differential growth of different crystal facets to produce complicated shapes.

As with other proteins, outside the crystallization region, depending on the rAAV serotype, various non-crystallization outcomes such as gel/oil phase, precipitate, and cluster are observed. The formation of dense solid phase/opaque crystals, which is not observed in low molecular weight or macromolecular protein crystallization (Trilisky et al., 2011), is probably a characteristic of crystallization of macromolecular assemblies of heavy molecular weight. Though the formation of well-defined drops/spots of dense phase/gel phase and their subsequent conversion into crystals are common in protein crystallization (Liu et al., 2010; Trilisky et al., 2011), the sluggishness in conversion of gel/oil spots into solids (7–12 days) for capsids compared to proteins is probably a characteristic of the crystallization of macromolecular assemblies of massive molecular weight and bulky nature, which significantly reduces their mobility in the dense phase to be transported either to form nuclei or to be attached to the nucleus. Because of the similar capsid structure and outer surface morphology, with spherical shape with T=1 icosahedral symmetry and 2-fold, 3-fold, and 5-fold symmetry, both full and empty capsids are expected to have a similar crystallization behavior but, in reality, they show completely different solution crystallization and non-crystallization behavior, which is attributed to the difference in their PTMS level and the charge environment surrounding them as full capsids carry DNA with strong negative charges in addition to the capsid's own surface charge.

Another interesting finding, which is not common in low molecular weight proteins (Abe et al., 2022; Iwai et al., 2008; Trilisky et al., 2011) and probably specific to the proteinaceous macromolecular assemblies with massive molecular weight, is that the extent of modification in their functional groups (i.e., PTMS level) can be so high that it can alter the trend in crystallization as observed by the completely reverse trend in crystallization of full and empty capsids between serotypes rAAV5 and rAAV9, and rAAV8.

**Author contributions:**
V.B. conceptualized this work, designed and conducted the experiments, analyzed the data, and wrote the initial draft. R.D.B. conceptualized this work, supervised the work/project, edited the manuscript, and acquired the funds to support the project. R.M.K., P.W.B., and S.L.S. edited the manuscript, supervised the work/ project and acquired the funds to support the project. A.J.S. and J.M.W. edited the manuscript, supervised the project and acquired the funds to support the project.

**Acknowledgements:**
Funding is acknowledged from the MLSC, Sanofi, Sartorius, Artemis, and US FDA (75F40121C00131).

# References

Abe, S., Tanaka, J., Kojima, M., Kanamaru, S., Hirata, K., Yamashita, K., Kobayashi, A., Ueno, T., 2022. Cell-free protein crystallization for nanocrystal structure determination. Sci Rep 12, 16031.

Ahamed, T., Esteban, B.N.A., Ottens, M., van Dedem, G.W.K., van der Wielen, L.A.M., Bisschops, M.A.T., Lee, A., Pham, C., Thömmesy, J., 2007. Phase behavior of an intact monoclonal antibody. Biophys J 93, 610–619.

Arakawa, T., Timasheff, S.N., 1985. Theory of protein solubility. Methods Enzymol. 114, 49–77.

Gilles, R., Hoffmann, E.K., Bolis, L., Arawaka, T., 2011. Advances in comparative and environmental physiology: Volume and osmolality control in animal cells. Springer-Verlag.

Asherie, N., 2004. Protein crystallization and phase diagrams. Methods 34, 266–272. https://doi.org/10.1016/j.ymeth.2004.03.028

Bal, V., Hong, M.S., Wolfrum, J.M., Barone, P.W., Springs, S.L., Sinskey, A.J., Kotin, R.M., Braatz, R.D., 2024a. An integrated experimental and modeling approach for crystallization of complex biotherapeutics. https://doi.org/arXiv.2412.09821

Bal, V., Wolfrum, J.M., Barone, P.W., Springs, S.L., Sinskey, A.J., Kotin, R.M., Braatz, R.D., 2024b. Selective Enrichment of Full AAV Capsids. https://doi.org/arXiv:2412.06093

Bharmoria, P., Gupta, H., Mohandas, V.P., Ghosh, P.K., Kumar, A., 2012. Temperature invariance of NaCl solubility in water: Inferences from salt-water cluster behavior of NaCl, KCl, and NH4Cl. Journal of Physical Chemistry B 116, 11712–11719. https://doi.org/10.1021/jp307261g

Bonneté, F., 2007. Colloidal approach analysis of the marseille protein crystallization database for protein crystallization strategies. Cryst Growth Des 7, 2176–2181. https://doi.org/10.1021/cg700711a

Broide, M.L., Berland, C.R., Pande, J., Ogun, O.O., Benedek, G.B., 1991. Binary-liquid phase separation of lens protein solutions. Proc Natl Acad Sci U S A 88, 5660–5664. https://doi.org/10.1073/pnas.88.13.5660

Brunchi, C.E., Ghimici, L., 2013. PEG in aqueous salt solutions. Viscosity and separation ability in a TiO2 suspension. Revue Roumaine de Chimie 58, 183–188. https://doi.org/10.13140/2.1.1574.4646

Charles, M., Veesler, S., Bonneté, F., 2006. MPCD: A new interactive on-line crystallization data bank for screening strategies. Acta Crystallogr D Biol Crystallogr 62, 1311–1318. https://doi.org/10.1107/S0907444906027594

Chernov, A.A., 2003. Protein crystals and their growth. J Struct Biol 142, 3–21. https://doi.org/10.1016/S1047-8477(03)00034-0

Cotmore, S.F., Tattersall, P., 2013. Parvovirus diversity and DNA damage responses. Cold Spring Harb Perspect Biol 5, a012989.

Curtis, R.A., Lue, L., 2006. A molecular approach to bioseparations: Protein-protein and protein-salt interactions. Chem Eng Sci 61, 907–923. https://doi.org/10.1016/j.ces.2005.04.007

Dumetz, A.C., Snellinger-O'Brien, A.M., Kaler, E.W., Lenhoff, A.M., 2007. Patterns of protein-protein interactions in salt solutions and implications for protein crystallization.

Protein Science 16, 1867–1877. https://doi.org/10.1110/ps.072957907

Dunbar, C.E., High, K.A., Joung, J.K., Kohn, D.B., Ozawa, K., Sadelain, M., 2018. Gene therapy comes of age. Science (1979) 359, 1–3.

Fattah, F.J., Lichter, N.F., Fattah, K.R., Oh, S., Hendrickson, E.A., 2008. Ku70, an essential gene, modulates the frequency of rAAV-mediated gene targeting in human somatic cells. Proc Natl Acad Sci USA 105, 8703–8708.

Ferré-D'Amaré, A.R., Zhou, K., Doudna, J.A., 1998. A general module for RNA crystallization. J Mol Biol 279, 621–631.

Ginn, S.L., Amaya, A.K., Alexander, I.E., Edelstein, M., Abedi, M.R., 2018. Gene therapy clinical trials worldwide to 2017: An update. Journal of Gene Medicine 20, 1–16.

Gordon, A.J. & Ford, F.A., 1972. The chemist's companion: A handbook of practical data, techniques and references, 1st ed. John Wiley & Sons, New York.

Grover, P.K., Ryall, R.L., 2005. Critical appraisal of salting-out and its implications for chemical and biological sciences. Chem Rev 105, 1–10. https://doi.org/10.1021/cr030454p

Gruntman, A.M., Su, L., Su, Q., Gao, G., Mueller, C., Flotte, T.R., 2015. Stability and compatibility of recombinant adeno-associated virus under conditions commonly encountered in human gene therapy trials. Hum Gene Ther Methods 26, 71–76. https://doi.org/10.1089/hgtb.2015.040

Haas, C., Drenth, J., Wilson, W.W., 1999. Relation between the solubility of proteins in aqueous solutions and the second virial coefficient of the solution. J. Phys. Chem. B 103, 2808–2811.

Halder, S., Vliet, K. Van, Smith, J.K., Duong, T.T.P., McKenna, R., Wilson, J.M., Agbandje-McKenna, M., 2015. Structure of Neurotropic Adeno-Associated Virus AA-Vrh.8. . J Struct Biol. 192, 21–36.

Heeb, R., Lee, S., Venkataraman, N. V., Spencer, N.D., 2009. Influence of salt on the aqueous lubrication properties of end-grafted, ethylene glycol-based self-assembled monolayers. ACS Appl Mater Interfaces 1, 1105–1112. https://doi.org/10.1021/am900062h

Hekmat, D., 2015. Large-scale crystallization of proteins for purification and formulation. Bioprocess Biosyst Eng 38, 1209–1231. https://doi.org/10.1007/s00449-015-1374-y

Hyde, A.M., Zultanski, S.L., Waldman, J.H., Zhong, Y.L., Shevlin, M., Peng, F., 2017. General principles and strategies for salting-out informed by the Hofmeister series. Org Process Res Dev 21, 1355–1370. https://doi.org/10.1021/acs.oprd.7b00197

Iwai, W., Yagi, D., Ishikawa, T., Ohnishi, Y., Tanaka, I., Niimura, N., 2008. Crystallization and evaluation of hen egg-white lysozyme crystals for protein pH titration in the crystalline state. J. Synchrotron Rad. 15, 312–315.

Jiang, J., Xiong, Y.L., Chen, J., 2010. pH Shifting Alters Solubility Characteristics and Thermal Stability of soy protein isolate and its globulin fractions in different pH, salt concentration, and temperature conditions. J. Agric. Food Chem. 58, 8035–8042.

Kaludov, N.E.P., Padron, E., Govindasamy, L., McKenna, R., Chiorini, J.A., Agbandje-McKenna, M., 2003. Production, purification and preliminary X-ray crystallographic studies of adeno-associated virus serotype 4. Virology 306, 1–6.

Krauss, I.R., Merlino, A., Sica, F., 2013. An overview of biological macromolecule crystallization. Int. J. Mol. Sci. 14, 11643–11691.

Kurth, S., Li, T., Hausker, A., Evans, W.E., Dabre, R., Müller, E., Kervinen, J., 2024. Separation of full and empty adeno-associated virus capsids by anion-exchange chromatography using choline-type salts. Anal Biochem 686, 115421.

Lerch, T.F., Xie, Q., Ongley, H.M., Hare, J., Chapman, M.S., 2009. Twinned crystals of adeno-associated virus serotype 3b prove suitable for structural studies. Acta Crystallogr Sect F Struct Biol Cryst Commun 65, 177–183.

Lins-Austin, B., Patel, S., Mietzsch, M., Brooke, D., Bennett, A., Venkatakrishnan, B., Van Vliet, K., Smith, A.N., Long, J.R., McKenna, R., Potter, M., Byrne, B., Boye, S.L., Bothner, B., Heilbronn, R., Agbandje-McKenna, M., 2020. Adeno-associated virus (AAV) capsid stability and liposome remodeling during endo/lysosomal pH trafficking. Viruses 12, 1–18. https://doi.org/10.3390/v12060668

Liu, Y., Wang, X., Ching, C.B., 2010. Toward further understanding of lysozyme crystallization: phase diagram, protein-protein interaction, nucleation kinetics, and growth kinetics. Cryst Growth Des 10, 548–558.

Ma, C.C., Wang, Z.L., Xu, T., He, Z.Y., Wei, Y.Q., 2020. The approved gene therapy drugs worldwide: from 1998 to 2019. Biotechnol Adv 40, 107502.

Martin, E.W., Harmon, T.S., Hopkins, J.B., Chakravarthy, S., Incicco, J.J., Schuck, P., Soranno, A., Mittag, T., 2021. A multi-step nucleation process determines the kinetics of prion-like domain phase separation. Nat Commun 12. https://doi.org/10.1038/s41467-021-24727-z

Mary, B., Maurya, S., Arumugam, S., Kumar, V., Jayandharan, G.R., 2019. Post-translational modifications in capsid proteins of recombinant adeno-associated virus (AAV) 1-rh10 serotypes. FEBS J 286, 4964–4981.

McGarvey, P.W., Hoffmann, M.M., 2018. Solubility of some mineral salts in polyethylene glycol and related surfactants. Tenside, Surfactants, Detergents 55, 203–209. https://doi.org/10.3139/113.110555

McPherson, A., Gavira, J.A., 2014. Introduction to protein crystallization. Acta Crystallogr F Struct Biol Commun 70, 2–20.

Mehta, Chirag M, White, E.T., Litster, J.D., 2012. Correlation of second virial coefficient with solubility for proteins in salt solutions. Biotechnol Prog 28, 163–170.

Mehta, Chirag M., White, E.T., Litster, J.D., 2012. Correlation of second virial coefficient with solubility for proteins in salt solutions. Biotechnol Prog 28, 163–170. https://doi.org/10.1002/btpr.724

Mikals, K., Nam, H.J., Vliet, K. Van, Vandenberghe, L.H., Mays, L.E., McKenna, R., Wilson, J.M., Agbandje-McKenna, M., 2014. The structure of AAVrh32.33, a novel gene delivery vector. J Struct Biol. 186, 308–317.

Miller, E.B., Whitaker, B.G., Govindasamy, L., McKenna, R., Zolotukhin, S., Muzyczka, N., McKenna, M.A., 2006. Production, purification and preliminary X-ray crystallographic studies of adeno-associated virus serotype 1. Acta Crystallogr Sect F Struct Biol Commun 62, 1271–1274.

Mitchell, H.M., Jovannus, D., Rosbottom, I., Link, F.J., Mitchell, N.A., Heng, J.Y.Y., 2023. Process modelling of protein

crystallization: A case study of lysozyme. Chemical Engineering Research and Design 192, 268–279.

Mitchell, M., Nam, H.J., Carter, A., McCall, A., Rence, C., Bennett, A., Gurda, B., McKenna, R., Porter, M., Sakai, Y., Byrne, B.J., Muzyczka, N., Aslanidi, G., Zolotukhin, S., Agbandje McKennaa, M., 2009. Production, purification and preliminary X-ray crystallographic studies of adeno-associated virus serotype 9. Acta Crystallogr Sect F Struct Biol cryst Commun 65, 715–718.

Nam, H.J., Gurda, B.L., McKenna, R., Potter, M., Byrne, B., Salganik, M., Muzyczka, N., Agbandje McKenna, M., 2011. Structural Studies of Adeno-Associated Virus Serotype 8 Capsid Transitions Associated with Endosomal Trafficking. J Virol 85, 11791–11799.

Nam, H.J., Lane, M.D., Padron, E., Gurda, B., McKenna, R., Kohlbrenner, E., Aslanidi, G., Byrne, B., Muzyczka, N., Zolotukhin, S., Agbandje McKenna, M., 2007. Structure of Adeno-Associated Virus Serotype 8, a Gene Therapy Vector. J Virol 81, 12260–12271.

Ozbas, B., Kretsinger, J., Rajagopal, K., Schneider, J.P., Pochan, D.J., 2004. Salt-triggered peptide folding and consequent self-assembly into hydrogels with tunable modulus. Macromolecules 37, 7331–7337. https://doi.org/10.1021/ma0491762

Pegram, L.M., Wendorff, T., Erdmann, R., Shkel, I., Bellissimo, D., Felitsky, D.J., Record, M.T.Jr., 2010. Why Hofmeister effects of many salts favor protein folding but not DNA helix formation. Proc Natl Acad Sci USA 107, 7716–7721.

Quigley, A., Williams, D.R., 2015a. The second virial coefficient as a predictor of protein aggregation propensity: A self-interaction chromatography study. European Journal of Pharmaceutics and Biopharmaceutics 96, 282–290.

Quigley, A., Williams, D.R., 2015b. The second virial coefficient as a predictor of protein aggregation propensity: A self-interaction chromatography study. European Journal of Pharmaceutics and Biopharmaceutics 96, 282–290. https://doi.org/10.1016/j.ejpb.2015.07.025

Rakel, N., Baum, M., Hubbuch, J., 2014. Moving through three-dimensional phase diagrams of monoclonal antibodies. Biotechnol. Prog. 13, 1103.

Randolph, A. D.; Larson, M.A., 1971. Theory of Particulate Processes. Elsevier, Amsterdam.

Rasubala, L., Fourmy, D., Ose, T., Kohda, D., Maenaka, K., Yoshizawa, s., 2005. Crystallization and preliminary X-ray analysis of the mRNA-binding domain of elongation factor SelB in complex with RNA. Acta Crystallogr Sect F: Struct Biol Cryst Commun. 61, 296–298.

Saikumar, M. V., Glatz, C.E., Larson, M.A., 1998. Lysozyme crystal growth and nucleation kinetics. J Cryst Growth 187, 277–288.

Schmittschmitt, J., Scholtz, P.J.M., 2003. The role of protein stability, solubility, and net charge in amyloid fibril formation. Protein Science 12, 2374–2378.

Sedgwick, H., Kroy, K., Salonen, A., Robertson, M.B., Egelhaaf, S.U., Poon, W.C.K., n.d. Non-equilibrium behavior of lysozyme solutions: beads, clusters and gels. https://arxiv.org/pdf/cond-mat/0309616.

Shaw, K.L., Grimsley, G.R., Yakovlev, G.I., Makarov, A.A., Pace, C.N., 2001. The effect of net charge on the solubility, activity,

and stability of ribonuclease Sa. Protein Science 10, 1206–1215.

Soleymani, J., Jouyban-Gharamaleki, V., Kenndler, E., Jouyban, A., 2020. Measurement and modeling of sodium chloride solubility in binary mixtures of water + polyethylene glycol 400 at various temperatures. J Mol Liq 316, 113777. https://doi.org/10.1016/j.molliq.2020.113777

Sommer, J.M., Smith, P.H., Parthasarathy, S., Isaacs, J., Vijay, S., Kieran, J., Powell, S.K., McClelland, A., Wright, J.F., 2003. Quantification of adeno-associated virus particles and empty capsids by optical density measurement. Molecular Therapy 7, 122–128.

Trilisky, E., Gillespie, R., Osslund, T.D., Vunnum, S., 2011. Crystallization and liquid-liquid phase separation of monoclonal antibodies and fc-fusion proteins: Screening results. Biotechnol Prog 27, 1054–1067.

Venkatakrishnan, B., Yarbrough, J., Domsic, J., Bennett, A., Bothner, B., Kozyreva, O.G., Samulski, R.J., Muzyczka, N., McKenna, R., McKenna, M.A., 2013. Structure and dynamics of adeno-associated virus serotype 1 VP1-unique N-terminal domain and its role in capsid trafficking. J Virol Methods 87, 4974–4984.

Wagner, C., Fuchsberger, F.F., Innthaler, B., Lemmerer, M., Birner-Gruenberger, R., 2023. Quantification of empty, partially filled and full adeno-associated virus vectors using mass photometry. Int J Mol Sci. 24, 11033.

Weber, M., Jones, M.J., Ulrich, J., 2008. Crystallization as a purification method for jack bean urease: On the suitability of poly (ethylene glycol), Li2SO4, and NaCl as precipitants. Cryst Growth Des 8, 711–716.

Wörner, T.P., Bennett, A., Habka, S., Snijder, J., Friese, O., Powers, T., Agbandje McKenna, M., Heck, A.J.R., 2021. Adeno-associated virus capsid assembly is divergent and stochastic. Nat Commun 12, 1642.

Xie, Q., Bu, W., Bhatia, S., Hare, J., Somasundaram, T., Azzi, A., Chapman, M.S., 2002. The atomic structure of adeno-associated virus (AAV-2), a vector for human gene therapy. Proc Natl Acad Sci USA 99, 10405–10410.

Xie, Q., Hare, J., Turnigan, J., Chapman, M.S., 2004. Large-scale production, purification and crystallization of wild-type adeno-associated virus-2. J Virol Methods 122, 17–27.

Xie, Q., Ongley, H.M., Hare, J., Chapmana, M.S., 2008. Crystallization and preliminary X-ray structural studies of adeno-associated virus serotype 6. Acta Crystallogr Sect F Struct Biol Cryst Commun 64, 1074–1078.

Yamanaka, M., Inaka, K., Furubayashi, N., Matsushima, M., Takahashi, S., Tanaka, H., Sano, S., Sato, M., Kobayashi, T., Tanaka, T., 2011. Optimization of salt concentration in PEG-based crystallization solutions. J Synchrotron Radiat 18, 84–87. https://doi.org/10.1107/S0909049510035995

Zayas, J.F., 1997. Functionality of proteins in food, 1st ed. Springer-Verlag, Berlin.

Zhang, L., Tan, H., Fesinmeyer, R., Li, Catrone, D., Le, D., Remmele, R. & Zhang, J., 2012. Antibody solubility behavior in monovalent salt solutions reveals specific anion effects at low ionic strength. J. Pharm. Sci. 101, 965–977.

Ziegler, G.R., Foegeding, E.A., 1990. The gelation of proteins. Adv Food Nutr Res 34, 203–

298. https://doi.org/10.1016/S1043-4526(08)60008-X

# A Phase Diagram for Crystallization of a Complex Macromolecular Assembly

Vivekananda Bal[≠], Jacqueline M. Wolfrum[§], Paul W. Barone[§], Stacy L. Springs[§],
Anthony J. Sinskey[§ς], Robert M. Kotin[§]*, and Richard D. Braatz[≠§]

[≠]Department of Chemical Engineering, Massachusetts Institute of Technology, Cambridge, MA, USA
[§]Center for Biomedical Innovation, Massachusetts Institute of Technology, Cambridge, MA, USA
[ς]Department of Biology, Massachusetts Institute of Technology, Cambridge, MA, USA
*Gene Therapy Center, University of Massachusetts Chan Medical School, Worcester, MA, USA

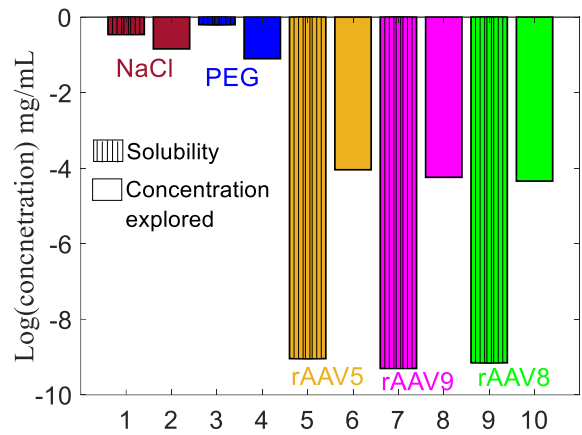

**Fig. S1:** Comparison of solubilities of NaCl, PEG8000, and rAAV with their concentrations used in crystallization process.

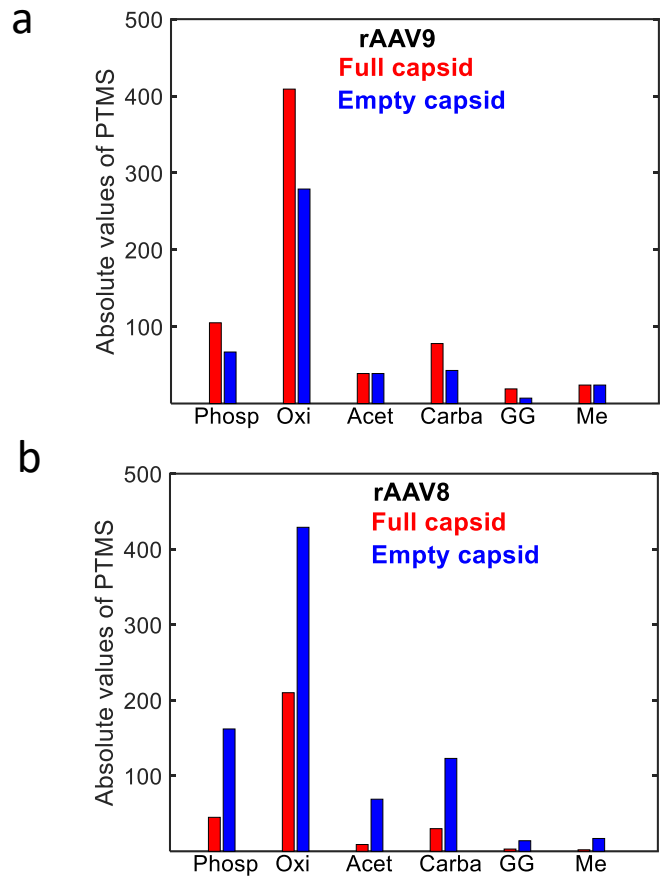

**Fig. S2:** Absolute values of post translational modifications (PTMS) in capsid proteins. (a) Comparison of PTMS values between full and empty capsids of sf9 produced rAAV9 capsids, (b) Comparison of PTMS values between sf9 and HEK produced rAAV8 capsids